\documentclass[10pt,conference, a4paper]{IEEEtran}
\usepackage{amsmath,amssymb}
\usepackage{graphicx}
\usepackage{mathrsfs}
\usepackage{tikz}
\usepackage{booktabs}
\usepackage{enumerate}
\usepackage{psfrag}	
\usepackage{array}
\usepackage{multirow,hhline}
\usepackage{exscale}
\usepackage{color}
\usepackage{colortbl}
\usepackage{epsfig,subfigure}
\usepackage{cite}
\usepackage{amsthm}
\newcommand{\mat}[1]{\ensuremath{\mathbf{#1}}}
\newcommand{\alert}[1]{\textcolor{black}{#1}}
\renewcommand{\vec}[1]{\ensuremath{\mathbf{#1}}}
\newtheorem{theorem}{Theorem}

\newtheorem{remark}{Remark}
\newtheorem{corollary}{Corollary}

\newcommand{\y}[0]{\ensuremath{\mathbf{y}}}
\newcommand{\z}[0]{\ensuremath{\mathbf{z}}}
\renewcommand{\H}[0]{\ensuremath{\mathbf{H}}}
\newcommand{\x}[0]{\ensuremath{\mathbf{x}}}
\newcommand{\X}[0]{\ensuremath{\mathbf{X}}}
\newcommand{\I}[0]{\ensuremath{\mathbf{I}}}
\newcommand{\D}[0]{\ensuremath{\mathbf{D}}}
\renewcommand{\u}[0]{\ensuremath{\mathbf{u}}}
\renewcommand{\v}[0]{\ensuremath{\mathbf{v}}}
\newcommand{\h}[0]{\ensuremath{\mathbf{h}}}
\renewcommand{\d}[0]{\ensuremath{\mathbf{d}}}
\newcommand{\w}[0]{\ensuremath{\mathbf{w}}}

\DeclareMathOperator{\tr}{trace}

\begin{document}

% \pagenumbering{none}

\IEEEoverridecommandlockouts
\title{Simultaneous Diagonalization: On the DoF Region of the K-user MIMO Multi-way Relay Channel}
\author{
\IEEEauthorblockN{Anas Chaaban, Karlheinz Ochs, and Aydin Sezgin}
\IEEEauthorblockA{Institute of Digital Communication Systems\\
Ruhr-Universit\"at Bochum (RUB), Germany\\
Email: {anas.chaaban,karlheinz.ochs,aydin.sezgin@rub.de}}
\thanks{%
This work is supported by the German Research Foundation, Deutsche
Forschungsgemeinschaft (DFG), Germany, under grant SE 1697/5.
}
}

\maketitle

% \footnotetext[1]{a}
%\thispagestyle{empty}

\begin{abstract}
The $K$-user MIMO Y-channel consisting of $K$ users which want to exchange messages among each other via a common relay node is studied in this paper. A transmission strategy based on channel diagonalization using zero-forcing beam-forming is proposed. This strategy is then combined with signal-space alignment for network-coding, and the achievable degrees-of-freedom region is derived. A new degrees-of-freedom outer bound is also derived and it is shown that the proposed strategy achieves this outer bound if the users have more antennas than the relay.
\end{abstract}

%\begin{IEEEkeywords}
%Multi-way relaying, DoF region, 
%\end{IEEEkeywords}

\section{Introduction}
Multi-way relaying has witnessed increasing research attention recently due to its importance from both theoretical and practical points of view. From a theoretical point of view, {the importance of multi-way relaying lies in being an environment where the potential of modern physical-layer network-coding schemes can be examined. From a practical point of view, its importance comes from modelling simultaneous communication between several base-stations via a common satellite for instance.}

The study of multi-way relaying started with the most fundamental case, the so-called two-way relay channel TWRC consisting of two users that want to exchange information through a common relay node. The TWRC has been studied thoroughly recently \cite{KimDevroyeMitranTarokh, GunduzTuncelNayak, AvestimehrSezginTse,  NamChungLee_IT, AlsharoaGhazzaiAlouini, ShaqfehZafarAlnuweiriAlouini} where several transmission strategies including compress-forward and lattice coding have been examined, leading to the capacity of the TWRC within a constant gap.

Several extensions of the TWRC have been considered lately. For instance, the multi-way relay channel (MWRC) with multi-cast messages \cite{MokhtarMohassebNafieElGamal, OngKellettJohnson_IT}, and the multi-pair TWRC \cite{SezginAvestimehrKhajehnejadHassibi, SezginBocheAvestimehr}. The Y-channel falls within this class of MWRC's. Namely, the Y-channel is a MWRC with full message exchange. It consists of 3-users that want to simultaneously exchange information in all directions via a relay. This channel has been studied in \cite{ChaabanSezginISIT, ChaabanSezgin_ISIT12_Y, ChaabanSezginAvestimehr_YC_SC} where the capacity of the linear-deterministic \cite{AvestimehrDiggaviTse_IT} case was determined, and the capacity of the Gaussian case a within a constant gap was characterized. The $K$-user Y-channel is similar to the 3-user case, but with more users communicating via a common relay. The capacity of the linear-deterministic 4-user Y-channel was characterized in \cite{ZewailMohassebNafieElGamal}.

The main focus of this paper is on the MIMO Y-channel. This has been introduced in \cite{LeeLimChun} where the strategy of signal-space alignment for network coding was used to characterize the degrees-of-freedom (DoF) of the 3-user MIMO Y-channel under some conditions on the number of antennas. However, a complete DoF characterization of the general 3-user MIMO Y-channel was not available until~\cite{ChaabanOchsSezgin} where a novel upper bound and a general transmission strategy were developed, thus settling this problem. The MIMO Y-channel with more than 3 users has also been studied in \cite{TianYenerMIMOMW,LeeLeeLee}. In \cite{TianYenerMIMOMW}, {Tian and Yener} have studied the multi-cluster MIMO Y-channel and characterized the DoF of the channel under some conditions on the number of antennas, while in \cite{LeeLeeLee}, {Lee {\it et al.}} proposed a transmission strategy for the $K$-user MIMO Y-channel and derived its achievable DoF. \alert{Recently, Wang has characterized the sum-DoF of the 4-user MIMO Y-channel in \cite{ChenweiWang}}. Despite the intensive work in this direction, the DoF of the $K$-user MIMO Y-channel is still an open problem. 

In this paper, we address this problem from the perspective of channel diagonalization. Namely, we propose a transmission strategy for the $K$-user MIMO Y-channel which is based on the simple strategy of zero-forcing beam-forming~\cite{Jindal}. {Namely, the beam-forming is realized by using a normalized version of the Moore-Penrose pseudo-inverse. This strategy leads to a simultaneous diagonalization of all uplink/downlink channels, thus revealing the structure of the optimal signal-space alignment for network-coding at the relay. By taking advantage of this insight, we derive the achievable DoF region of the proposed strategy.} We also derive a DoF outer bound based on genie-aided arguments, and then prove that our strategy achieves the outer bound as long as the users have more antennas than the relay. This provides {\it the first DoF region characterization for the $K$-user MIMO Y-channel.} 

{Although a complete characterization of the DoF of the $K$-user MIMO Y-channel is not available to date, the proposed strategy significantly simplifies the treatment of the channel. Furthermore, the authors believe that the insights gained from the channel diagonalization strategy in this paper will ease the way to completing the DoF characterization.}

The paper is organized as follows. We start by formally defining the $K$-user MIMO Y-channel in Section \ref{Sec:Model}. Then, we introduce the normalized pseudo-inverse in Section \ref{NMPPI} which is necessary for zero-forcing beam-forming. Next, we describe our transmission strategy in Section \ref{TransmissionStrategy}, and prove its optimality in Section \ref{DoFRegion}. Finally, we conclude the paper with a discussion in Section \ref{Discussion}. Throughout the paper, we use $\mathcal{CN}(\vec{m},\mat{Q})$ to denote a complex Gaussian random vector with mean $\vec{m}$ and covariance matrix $\mat{Q}$. We use $\I_N$ to denote the identity matrix of size $N\times N$ and $\vec{0}_q$ to denote a zero vector of length $q$. We use bold-face lower-case and upper-case letters to denote vectors and matrices, respectively. We use $\X^{H}$ and $\X^{-1}$ to denote the Hermitian transpose and the inverse of a matrix $\X$, respectively. We also use $\x^{n}$ to denote the length-$n$ sequence $(\x(1),\cdots,\x(n))$.

\section{System Model}
\label{Sec:Model}
\begin{figure}
\centering
\includegraphics[width=\columnwidth]{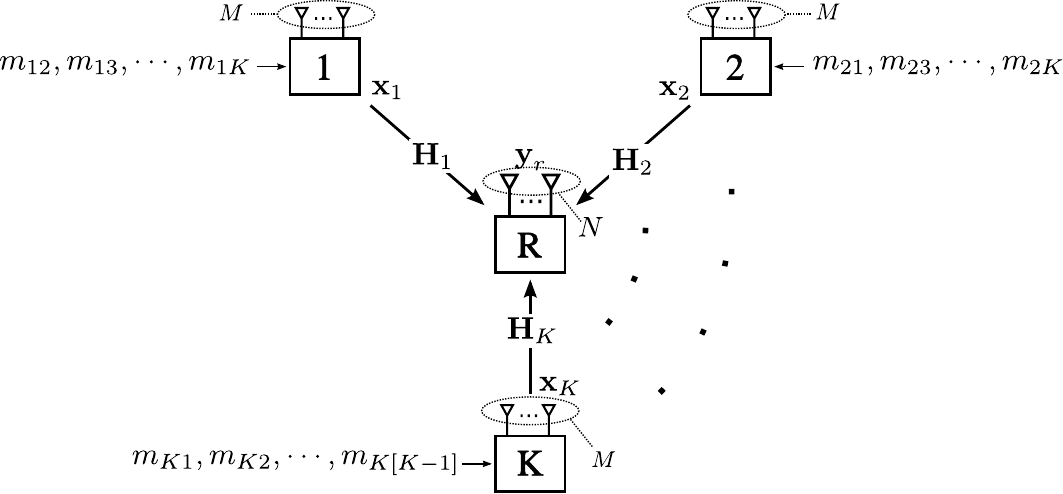}
\caption{The $K$-user MIMO Y-channel in the uplink. Each user $j\in\{1,\cdots,K\}$ sends a message $m_{jk}$ to each user $k\in\{1,\cdots,K\}$, $k\neq j$.}
\label{Fig:ModelU}
\end{figure}

\begin{figure}
\centering
\includegraphics[width=\columnwidth]{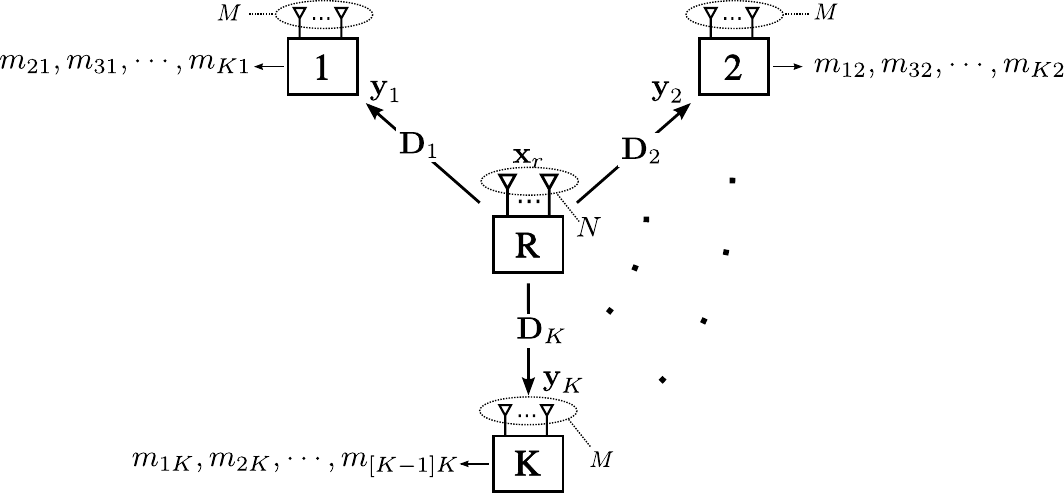}
\caption{The $K$-user MIMO Y-channel in the downlink. Each user $k\in\{1,\cdots,K\}$ decodes a message $m_{jk}$ from each user $j\in\{1,\cdots,K\}$, $k\neq j$.}
\label{Fig:ModelD}
\end{figure}

The $K$-user MIMO Y-channel consists of $K$ users which want to establish full message-exchange via a relay as shown in Fig. \ref{Fig:ModelU} and \ref{Fig:ModelD}. All nodes are assumed to be full-duplex\footnote{\alert{The results of the paper can be easily extended to the half-duplex case.}} with power $P$, the relay has $N$ antennas, and each user has $M$ antennas. User $j\in\{1,\cdots,K\}$ has a message $m_{jk}$ to be sent to user $k$ for all $k\in\{1,\cdots,K\}$ with $j\neq k$. The rate of message $M_{jk}$ is $R_{jk}(P)$. 

User $j$ sends the signal $\x_{j}(i)$, an $M\times 1$ complex-valued vector,  at time instant\footnote{The time index $i$ will be dropped in the sequel unless necessary} $i$. The received signal at the relay is given by (cf. Fig. \ref{Fig:ModelU})
\begin{align}
\y_{r}(i)=\sum_{j=1}^K\H_j\x_j(i)+\z_{r}(i),
\end{align}
which is an $N\times1$ vector, where the noise $\z_{r}(i)$ is $\mathcal{CN}(\vec{0},\vec{I})$ and is i.i.d. over time. Here $\H_j$ is the $N\times M$ complex channel matrix from user $j$ to the relay, which is assumed to \alert{be block-constant}. The relay transmit signal $\x_{r}(i)$ is an $N\times 1$ vector. The received signal at user $j$ is given by (cf. Fig. \ref{Fig:ModelD})
\begin{align}
\label{ReceivedSignal}
\y_{j}(i)=\D_j\x_{r}(i)+\z_{j}(i),
\end{align}
which is an $M\times1$ vector, where the noise $\z_{j}(i)$ is $\mathcal{CN}(\vec{0},\vec{I})$ and is i.i.d. over time, and $\D_j$ is the \alert{block-constant} $M\times N$ downlink complex channel matrix from the relay to user $j$. The transmit signals of the users and the relay must satisfy the power constraint, i.e., 
\begin{align}
\tr(\mathbb{E}[\x_j\x_j^{H}])&\leq P,\\ 
\tr(\mathbb{E}[\x_r\x_r^{H}])&\leq P.
\end{align}

The achievable rates and the capacity region of the MIMO Y-channel are defined in the standard information theoretic sense~\cite{CoverThomas}. Since we are interested in the DoF region of the channel, we define the the DoF of message $m_{jk}$ as 
\begin{align}
\label{DoFDef}
d_{jk}=\lim_{\substack{P\to\infty}}\frac{R_{jk}(P)}{\log(P)}.
\end{align}
A DoF $d_{jk}$ is said to be achievable if there exists a an achievable rate $R_{jk}(P)$ satisfying \eqref{DoFDef}. A DoF vector $\d\in\mathbb{R}^{K(K-1)}$ defined as
\begin{align*}
&\d=\\
&(d_{12},\cdots,d_{1K},d_{21},d_{23},\cdots,d_{2K},\cdots,d_{K1},\cdots,d_{K[K-1]}),\nonumber
\end{align*}
is said to be achievable if its components are simultaneously achievable. We define the DoF region as the set of all achievable vectors $\d$.

\section{Normalized Pseudo-inverse}
\label{NMPPI}

An essential quantity for the zero-forcing beam-forming strategy is the normalized Moore-Penrose pseudo-inverse (MPPI). The right-MPPI is defined for a matrix $\H_j\in\mathbb{C}^{N\times M}$ with $N\leq M$ as \cite{Bernstein}
\begin{align}
\H_j^{\dagger}=\H_j^{H}[\H_j\H_j^{H}]^{-1},
\end{align}
which is an $M\times N$ matrix satisfying
\begin{align}
\H_j\H_j^{\dagger}=\I_N.
\end{align}

The right-MPPI is useful for designing pre-coding matrices that diagonalize the channel. For instance, user $j$ can send $\x_j=\H_j^{\dagger}\u_j$ where $\u_j\in\mathbb{C}^{N\times 1}$ is a codeword symbol, which makes the received signal at the relay $\H_j\H_j^{\dagger}\u_j+\z_r=\u_j+\z_r$, thus achieving channel diagonalization. 

However, this pre-coding is allowed only if the power constraint is satisfied at the transmitter. This requires a normalized MPPI which does not change the power of $\u_j$, i.e., $$\tr(\mathbb{E}[\x_j\x_j^{H}])=\tr(\mathbb{E}[\u_j\u_j^{H}]).$$ For this purpose, we define the normalized right-MPPI as
\begin{align}
\H_j^{R}=\alpha_j\H_j^{\dagger},
\end{align}
with
\begin{align}
\alpha_j^{-2}=\tr([\H_j^{\dagger}]^{H}\H_j^{\dagger}).
\end{align}
This guarantees that $\x_j$ has the same power as $\u_j$. Notice that this leads to 
\begin{align}
\label{UplinkDiagonalization}
\H_j\H_j^{R}=\alpha_j\I_N
\end{align}
Similarly, we define the normalized left-MPPI for a matrix $\D_j$ with $M\times N$ dimensions satisfying $N\leq M$ as 
\begin{align}
\D_j^{L}=\beta_j\D_j^{\ddagger},
\end{align}
with 
\begin{align}
\D_j^{\ddagger}&=[\D_j^{H}\D_j]^{-1}\D_j^{H}\\ \beta_j^{-2}&=\tr([\D_j^{\ddagger}]^{H}\D_j^{\ddagger}).
\end{align}
This leads to
\begin{align}
\D_j^{L}\D_j=\beta_j\I_N.
\end{align}
This can be used to diagonalize the downlink channel from the relay to the users. Next, we describe our transmission strategy for the MIMO Y-channel.

\section{Transmission Strategy}
\label{TransmissionStrategy}
Using the normalized MPPI, we can design a pre- and post-coding strategies at the users which simultaneously diagonalizes the channels to and from the relay. Notice that since the channels $\H_j$ are in general different, diagonalizing $\H_1$ by post-coding at the relay does not necessarily diagonalize $\H_2$ and vice versa. Therefore, this simultaneous diagonalization has to be done at the users. 

For simplicity of exposition, we describe the transmission strategy for the $K=4$ user case. The extension to $K>4$ users is trivial. We start with the uplink phase from the users to the relay.

\subsection{Uplink}
Let us consider user 1 whose transmit signal is $\x_1$ and channel to the relay is $\H_1$. This user has a symbol $\u_1\in\mathbb{C}^{N}$ with power $P$ to be transmitted to the relay. This symbol contains information for all users $j\neq 1$, and its construction will be explained later on after we describe the uplink. User 1 uses zero-forcing beam-forming to pre-code $\u_1$ by using the normalized right-MPPI, i.e., $\x_1=\H_1^{R}\u_1$. 
\begin{remark}
{This is equivalent to sending the $i$th component $u_{1i}$ of $\u_1$ along a beam-forming vector $\h_{1i}^R$ which is the $i$th column of $\H_1^{R}$ which is orthogonal to all rows of $\H_1$ except the $i$th row.}
\end{remark}

Notice that since $\H_1^{R}$ is normalized, then $\x_1$ also has power $P$ and hence does not violate the power constraint.

Similarly, all users pre-code with their normalized right-MPPI, leading to
\begin{align}
\x_j=\H_j^{R}\u_j.
\end{align}
Now the received signal at the relay becomes
\begin{align}
\y_r&=\sum_{j=1}^{K}\H_j\x_j+\z_r\\
&=\sum_{j=1}^{K}\H_j\H_j^{R}\u_j+\z_r\\
&=\sum_{j=1}^{K}\alpha_j\u_j+\z_r
\end{align}
by \eqref{UplinkDiagonalization}. Now notice the desirable structure of $\y_r$. The relay simply obtains a scaled sum of all symbols $\u_j$. 

Denote the codeword symbol corresponding to communication from user $j$ to user $k\neq j$ as $\v_{jk}\in\mathbb{C}^{d_{jk}}$ for some $d_{jk}\in\mathbb{N}$. We construct $\u_j$ as a concatenation of the codeword symbols from user $j$ to all other users. If we construct $\u_j$ in such a way that the codeword symbols align pair-wise at the relay, then we obtain a network-code at the relay constructed by the channel (on-the-fly). To this end, define 
\begin{align}
\label{elljk}
\ell_{jk}=\ell_{kj}=\max\{d_{jk},d_{kj}\},
\end{align}
and define further $\u_{jk}$ as 
\begin{align}
\u_{jk}=\begin{bmatrix}\v_{jk}\\ \mathbf{0}_{\ell_{jk}-d_{jk}}\end{bmatrix}
\end{align}
The transmit signals of the 4 users are then constructed as follows 
\begin{figure}
\centering
\includegraphics[width=\columnwidth]{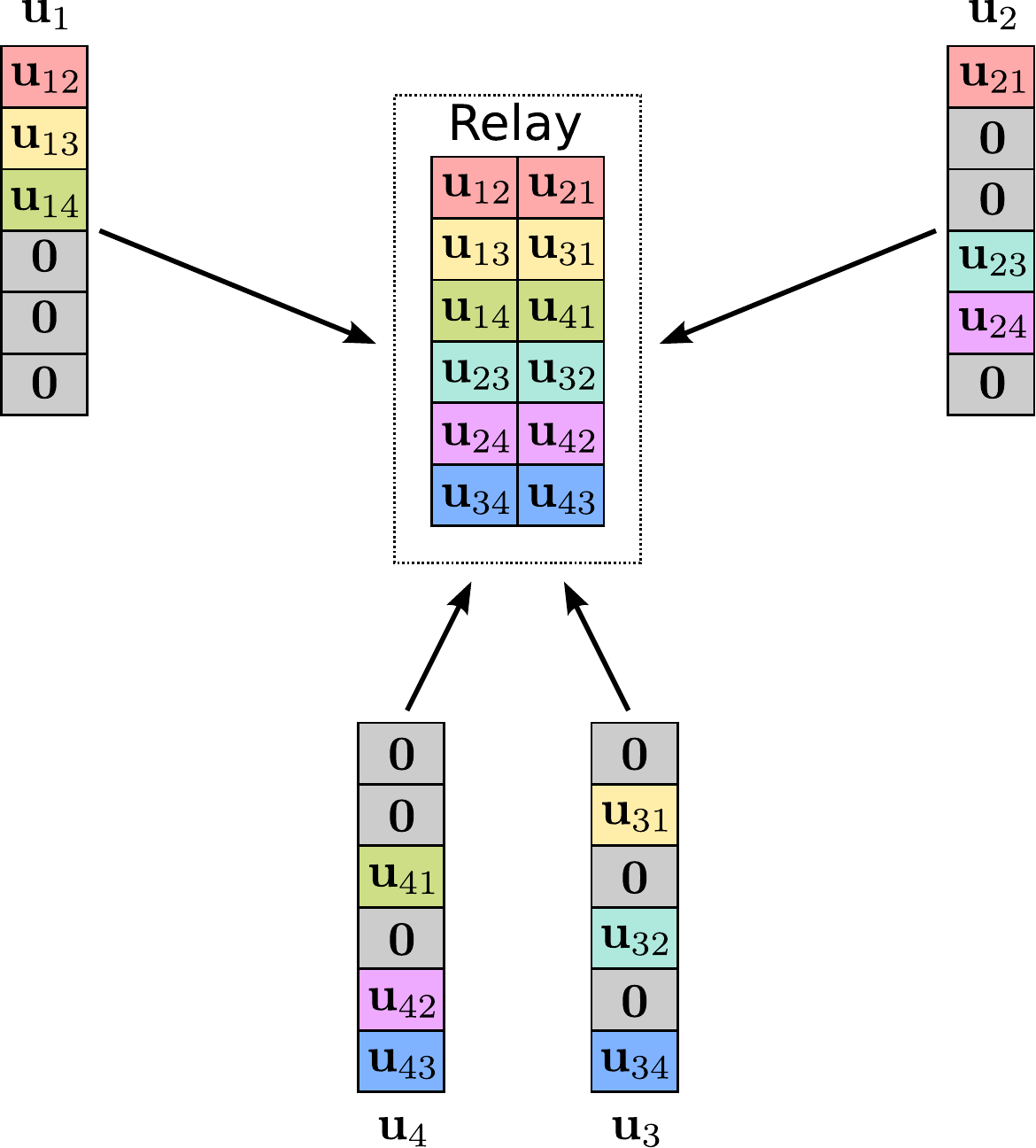}
\caption{A graphical illustration of the uplink phase in the 4-user MIMO Y-channel showing the pair-wise alignment of signals at the relay.}
\label{UplinkAlign}
\end{figure}
\begin{align}
\label{SignalUj}
\u_1=\begin{bmatrix} \u_{12}\\ \u_{13}\\ \u_{14}\\ \mathbf{0}_{\ell_{23}}\\ \mathbf{0}_{\ell_{24}}\\ \mathbf{0}_{\ell_{34}}\\ \mathbf{0}_{\ell_{0}}\end{bmatrix},\ 
\u_2=\begin{bmatrix} \u_{21}\\ \mathbf{0}_{\ell_{13}}\\ \mathbf{0}_{\ell_{14}}\\ \u_{23}\\ \u_{24}\\ \mathbf{0}_{\ell_{34}}\\ \mathbf{0}_{\ell_{0}}\end{bmatrix},\ 
\u_3=\begin{bmatrix} \mathbf{0}_{\ell_{12}}\\ \u_{31}\\ \mathbf{0}_{\ell_{14}}\\ \u_{32}\\ \mathbf{0}_{\ell_{24}}\\ \u_{34}\\ \mathbf{0}_{\ell_{0}}\end{bmatrix},\ 
\u_4=\begin{bmatrix} \mathbf{0}_{\ell_{12}}\\ \mathbf{0}_{\ell_{13}}\\ \u_{41}\\ \mathbf{0}_{\ell_{23}}\\ \u_{42}\\ \u_{43}\\ \mathbf{0}_{\ell_{0}}\end{bmatrix}
\end{align}
where {the last $\ell_0=N-\ell_{12}-\ell_{13}-\ell_{14}-\ell_{23}-\ell_{24}-\ell_{34}$ components are used to zero-pad the vectors to length $N$. Fig. \ref{UplinkAlign} shows a graphical illustration of the uplink which ignores the last $\ell_0$ components}. This choice of $\u_j$ establishes signal-space alignment for network-coding in the uplink \cite{NamChungLee_IT}. Using this construction, the received signal at the relay becomes
\begin{align}
\label{UserRelayChannel}
\y_r= \begin{bmatrix} \alpha_1\u_{12}+\alpha_2\u_{21}\\
\alpha_1\u_{13}+\alpha_3\u_{31}\\ 
\alpha_1\u_{14}+\alpha_4\u_{41}\\ 
\alpha_2\u_{23}+\alpha_3\u_{32}\\ 
\alpha_2\u_{24}+\alpha_4\u_{42}\\ 
\alpha_3\u_{34}+\alpha_4\u_{43}\\ 
\mathbf{0}_{\ell_{0}}\end{bmatrix}+\z_r=\w+\z_r.
\end{align}
Notice that this decomposes the channel to the relay into $N$ parallel SISO  TWRC's \cite{AvestimehrSezginTse}. Now the relay can decode all components of $\w$. This can be made possible by using nested-lattice codes for encoding the messages for instance \cite{NamChungLee_IT}. Using this construction, the relay obtains a network-code of each pair of symbols to be exchanged in a bi-directional manner ($j\leftrightarrow k$). Next, we explain the downlink phase.

\subsection{Downlink}
In the downlink phase, the relay forwards the decoded symbols to the users\footnote{The uplink and downlink phases take place simultaneously since all nodes are full-duplex.}. The relay transmit signal is a normalized version of $\w$ that satisfies the power constraint. Namely, 
\begin{align}
\x_r=\frac{\sqrt{P}}{\|\w\|}\w.
\end{align}
The received signal at user $k$ is then
\begin{align}
\y_k&=\D_k\x_r+\z_k\\
&=\frac{\sqrt{P}}{\|\w\|}\D_k\w+\z_k.
\end{align}
User $k$ then performs zero-forcing post-coding on $\y_k$ using the normalized left-MPPI which yields
\begin{align}
\label{RelayUserChannel}
\D_k^{L}\y_k=\frac{\sqrt{P}}{\|\w\|}\beta_k\w+\D_k^{L}\z_k.
\end{align}
Thus, after post-coding user $k$ observes $N$ parallel SISO point-to-point channels from the relay. Thus, user $k$ can decode $\w$, extract the desired components of $\w$, and then subtract the self interference to obtain the desired signals. For instance, user 1 decodes\footnote{\alert{The decodability of codeword sums can be enabled by using nested-lattice codes~\cite{NazerGastpar,ChaabanSezgin_IT_IRC} for instance.}} $\alpha_1\u_{12}+\alpha_2\u_{21}$, $\alpha_1\u_{13}+\alpha_3\u_{31}$, and $\alpha_1\u_{14}+\alpha_4\u_{41}$ and then subtracts its own self-interference to obtain $\u_{21}$, $\u_{31}$, and $\u_{41}$. Similarly, each user obtains his desired signals.

This completes the description of the transmission strategy. Next, we show that this strategy achieves the DoF region of the given network.

\section{DoF Region}
\label{DoFRegion}
The following theorem characterizes the DoF region of the 4-user case.

\begin{theorem}
\label{DoF4Users}
The DoF region $\mathcal{D}_4$ of the MIMO 4-user Y-channel with $N\leq M$ is given by the set of 12-tuples $\d\in\mathbb{R}^{12}_+$ satisfying 
\begin{align}
\label{DoFConstraint}
\sum_{j=1}^{4}\sum_{k=j+1}^{4}d_{p_jp_k}\leq N, \quad \forall \mathbf{p}
\end{align}
where $\mathbf{p}$ is a permutation of $(1,2,3,4)$ and $p_j$ is its $j$-th component.
\end{theorem}
The proof is provided in the next subsections. We start with the achievability.
\subsection{Achievability}
The achievability of this DoF region follows by examining the transmission strategy described above. Namely, it can be seen that the construction of the transmit signals $\u_j$ as given in \eqref{SignalUj} works if
\begin{align}
\label{AchievableDoFell}
\ell_{12}+\ell_{13}+\ell_{14}+\ell_{23}+\ell_{24}+\ell_{34}\leq N.
\end{align}
The channel to the relay is effectively a MIMO $N\times N$ point-to-point channel with input $\w$ and output $\y_r$ as given in \eqref{UserRelayChannel}. The DoF achievable over this channel is $N$. 

The users on the other hand can decode $\w$ since the channel from the relay to each user is effectively a MIMO $N\times N$ point-to-point channel as given in \eqref{RelayUserChannel}. The DoF achievable over this channel is $N$, and each user $k$ can obtain his desired signal as long as $d_{jk}\leq \ell_{jk}$ which is guaranteed by \eqref{elljk}.

Now, by substituting  \eqref{elljk} in the condition \eqref{AchievableDoFell}, we obtain \eqref{DoFConstraint}. However, there is still the restriction that $\ell_{jk}\in\mathbb{N}$ which might not allow achieving all DoF vectors in $\mathcal{D}_4$. To overcome this problem, we use symbol extensions as in \cite{ChaabanSezginISIT} to show the achievability of DoF vectors in $\mathbb{Q}_+^{12}\cap\mathcal{D}_4$ \alert{where $\mathbb{Q}$ is the set of rational numbers}. Since $\mathcal{D}_4$ is a polytope whose faces are defined by linear inequalities with integer coefficients \eqref{DoFConstraint}, then its corner points must be in $\mathbb{Q}_+^{12}$. Therefore, our strategy achieves all corner points of $\mathcal{D}_4$ and by time sharing, achieves all $\d\in\mathcal{D}_4$ which completes the proof of the achievability of the DoF region in Theorem \ref{DoF4Users}.

\subsection{Converse}
The converse of Theorem \ref{DoF4Users} can be shown by using the genie-aided upper bound in \cite{ChaabanSezginAvestimehr_YC_SC}. Let us consider $n$ uses of the channel, and let us give $(m_{23},m_{24},\y_2^{n})$ and $(m_{34},\y_3^{n})$ to user 1 as side information.

Consider any achievable rate for the channel, for which every node can obtain its messages with an arbitrarily small probability of error. This means that, after $n$ channel uses, user 1 can decode $m_{21}$, $m_{31}$, and $m_{41}$ from $(\y_1^{n}, m_{12}, m_{13}, m_{14})$. After decoding its desired messages, user 1 knows $(\y_2^{n}, m_{21}, m_{23}, m_{24})$. This makes user 1 able to decode $m_{32}$ and $m_{42}$ since user 2 can decode them from the same observation. After this step, user 1 has knowledge of $(\y_3^{n}, m_{31}, m_{32}, m_{34})$ which allows him to decode $m_{43}$ since user 3 can decode it. 

All in all, from the knowledge of $\y_1^{n}$, $\y_2^{n}$, $\y_3^{n}$ and $m_{12}$, $m_{13}$, $m_{14}$, $m_{23}$, $m_{24}$, and $m_{34}$, user 1 can decode all 6 remaining messages. Using Fano's inequality \cite{CoverThomas}, and defining $\widehat{M}_1=(M_{21},M_{31},M_{41})$ and $M_1=(M_{12},M_{13},M_{14})$, we can write\footnote{We drop the dependence of $R_{jk}$ on $P$ for clarity.}
\begin{align*}
&n(R_{21}+R_{31}+R_{41}+R_{32}+R_{42}+R_{43}-\varepsilon_n)\nonumber\\
&\leq I(\widehat{M}_{1},M_{32},M_{42},M_{43};\vec{y}_1^n,\vec{y}_2^n,\vec{y}_3^n,M_1,M_{23},M_{24},M_{34})\nonumber\\
&\leq h(\vec{y}_1^n,\vec{y}_2^n,\vec{y}_3^n)-h(\vec{y}_1^n,\vec{y}_2^n,\vec{y}_3^n|\x_r^{n})\\
&= I(\x_r^{n};\vec{y}_1^n,\vec{y}_2^n,\vec{y}_3^n)
\end{align*}
where $\varepsilon_n\to0$ as $n\to\infty$, and where the second step follows by using the definition of mutual information and the fact that conditioning does not increase entropy. We can write this bound as
\begin{align}
&n(R_{21}+R_{31}+R_{41}+R_{32}+R_{42}+R_{43}-\varepsilon_n)\nonumber\\
&\leq I(\x_r^{n};\D \x_r^{n}+\z^{n})
\end{align}
where
\begin{align}
\D=\begin{bmatrix}\D_1\\\D_2\\\D_3\end{bmatrix},\quad \text{and}\quad \z=\begin{bmatrix}\z_1\\\z_2\\\z_3\end{bmatrix}.
\end{align}
But this is the mutual information between the input $\x_r$ and the output $\D\x_r+\z$ of a MIMO $N\times 3M$ point-to-point channel. This channel has $\min\{N,3M\}=N$ DoF. Therefore, by dividing by $n$ and then letting $n\to\infty$ we get
\begin{align*}
R_{21}+R_{31}+R_{41}+R_{32}+R_{42}+R_{43}\leq  N\log(P) +n\mathcal{O}(1),
\end{align*}
which proves that 
\begin{align}
d_{21}+d_{31}+d_{41}+d_{32}+d_{42}+d_{43}\leq  N,
\end{align}
which is equivalent to \eqref{DoFConstraint} with $\vec{p}=(4,3,2,1)$. The upper bounds for all other permutations can be proved similarly.

This concludes the proof of the converse of Theorem \ref{DoF4Users} and shows the optimality of our diagonalization strategy for the 4-user MIMO Y-channel. This statement is further extended to the $K$-user case in the next section.

\section{Discussion}
\label{Discussion}
The proposed transmission strategy serves all users of the Y-channel and is able to support asymmetric DoF allocation between different streams. This makes it able to achieve the DoF region of the 4-user MIMO Y-channel with more antennas at the users than the relay. In fact, the same strategy can be used to achieve the DoF region of the $K$-user case given in the following corollary.
\begin{corollary}
\label{DoFKUsers}
The DoF region $\mathcal{D}_K$ of the MIMO $K$-user Y-channel with $N\leq M$ is given by the set of tuples $\d\in\mathbb{R}^{K(K-1)}$ satisfying 
\begin{align}
\sum_{j=1}^{K}\sum_{k=j+1}^{K}d_{p_jp_k}\leq N, \quad \forall \mathbf{p}
\end{align}
where $\mathbf{p}$ is a permutation of $(1,\cdots,K)$ and $p_j$ is its $j$-th component.
\end{corollary}
The corollary is an extension of Theorem \ref{DoF4Users} and can be proved similarly. {From this corollary, we can conclude that the sum-DoF is $2N$ if $N\leq M$ which coincides with the sum-DoF result in \cite{TianYenerMIMOMW} under this condition.}

Just as channel diagonalization is useful in the MIMO point-to-point channel \cite{Telatar}, it is also useful in other channels such as the Y-channel. It leads to a simpler representation of the channel, and can be used to design optimal transmission strategies. Our proposed strategy is an example of simultaneous channel diagonalization of several channels over a multi-user relay channel, leading to a simple structure of the relay received signal. The diagonalization is established using the simple scheme of zero-forcing beam-forming with the aid of the MPPI. The drawback of our strategy is its limitation to $N\leq M$ which is due to the use of the MPPI, which only exists under this condition. It would be interesting to develop simultaneous diagonalization strategies for such networks that overcome this limitation. For instance, one can use higher-order generalized SVD's or block SVD's to diagonalize the channels. It is worth at this point to mention that these two possibilities have been examined in \cite{AndreMasterThesis}.

One direction for future work is to further investigate diagonalization  strategies and other possibilities, with the aim of characterizing the DoF of the general $K$-user MIMO Y-channel, which is unknown so far.

\bibliography{myBib}

\end{document}